Generation
December 4-5, 1998.
6.0 document, and
conference format for 
\documentclass[final]{siamltex}
\usepackage [dvips]{graphicx}

\title{Hash Sort: A Linear Time Complexity Mulitiple-Dimensional Sort 
Algorithm \\ 
\emph{originally entitled "Making a Hash of Sorts"} 
\thanks{Special thanks and appreciation to Dr. Michael Mascagni for the 
opportunity to present and publish this paper.}}

\author{William F. Gilreath \\ ( {\tt will@williamgilreath.com} ) }

\date{December 1998}

\begin{document}

\maketitle

\vspace{3 mm}

\begin{abstract}

\vspace{3 mm}

\indent
Sorting and hashing are two completely different concepts in computer 
science, and appear mutually 
exclusive to one another. Hashing is a search method using the data as a 
key to map to the location within 
memory, and is used for rapid storage and retrieval.  Sorting is a 
process of organizing data from a random 
permutation into an ordered arrangement, and is a common activity 
performed frequently in a variety of 
applications.

\vspace{3 mm}

Almost all conventional sorting algorithms work by comparison, and in 
doing so have a linearithmic 
greatest lower bound on the algorithmic time complexity. Any improvement 
in the theoretical time complexity 
of a sorting algorithm can result in overall larger gains in 
implementation performance. A gain in algorithmic 
performance leads to much larger gains in speed for the application that 
uses the sort algorithm. Such a sort 
algorithm needs to use an alternative method for ordering the data than 
comparison, to exceed the linearithmic 
time complexity boundary on algorithmic performance.

\vspace{ 2 mm}

The hash sort is a general purpose non-comparison based sorting algorithm 
by hashing, which has some 
interesting features not found in conventional sorting algorithms. The 
hash sort asymptotically outperforms the 
fastest traditional sorting algorithm, the quick sort. The hash sort 
algorithm has a linear time complexity factor -- 
even in the worst case. The hash sort opens an area for further work and 
investigation into alternative means 
of sorting. 

\end{abstract}

\section{Theory}

\subsection{Sorting}

\indent\indent

Sorting is a common processing activity for computers to perform. Data 
that is sorted is of the form where 
data items have an increasing value -- ascending, or alternatively 
decreasing value -- descending. Regardless of 
the arrangement form, sorting establishes an ordering of the data. The 
arrangement of data from some random 
configuration into an ordered one is necessary often in many algorithms, 
applications, and programs. The need 
for sorting to arrange and order data makes the space and temporal 
complexity of the algorithm used very 
paramount. 

\vspace{3 mm}

A bad choice of algorithm for sorting data by a designer or programmer 
can result in mediocre 
performance in the end. With the large need for sorting, and the 
important concern of performance, many 
different types and kinds of algorithms for sorting have been devised. 
Some algorithms used such as quick sort 
and bubble sort are very widespread and often used. Other algorithms such 
as bin sort and pigeonhole sort are 
not as widely known. 

\vspace{3 mm}

The plethora of sorting algorithms available does not change the 
tantamount question of how fast it is 
possible to sort. This is a question of temporal efficiency, and is the 
most significant criterion for a sort 
algorithm. Along with it, what will affect the temporal efficiency and 
why it does is just as important a concern. 
Still another important concern, though more subtle, is the space 
requirements for using the algorithm -- space 
efficiency. This is a matter of algorithm overhead and resource 
requirements involved in the sorting process. 

\vspace{3 mm}

All sort algorithms for the most part have the temporal efficiency and 
causes for changes in it are well 
understood. This limit or barrier to sorting speed is a greatest lower 
bound which is $O(N\  log\ N )$. In essence, 
the reasoning  behind this is that sorting is a comparative or decision 
making process of $N$ items, which forms a 
binary tree of depth $log\ N$. For each data item, a decision must be 
made where to move it to maintain the 
ordering property desired. With $N$ data items, and $log\ N$ decision 
time, the minimum speed possible is the 
product, which is $N\ log\ N$. This lower bound is often never reached in 
practice, it is often a multiplied or added 
constant away from the theoretical maximum. 

\vspace{3 mm}

There is no theoretic greatest lower bound for space efficiency, and 
often this complexity measure 
characterizes a sort algorithm. The space requirements are highly 
dependent on the underlying method used in 
the sort algorithm, so space efficiency directly reflects this. While no 
theoretical limit is available, an optimum 
sort algorithm will require $N + 1$ storage. This is the most optimal 
because $N$ data items, and one additional data 
item used by the sort algorithm. A bubble sort algorithm has this optimal 
storage efficiency.  However, the 
optimal space efficiency is subordinate to the temporal efficiency in a 
sorting algorithm. The bubble sort, while 
having an optimal space efficiency, is well reputed to be very poor in 
performance, far above the theoretical 
lower bound. 
	
\subsection{Hashing}

\indent\indent

Hashing is a process of searching through data, using each data item 
itself as a key in the search. Hashing 
does not order the data items with respect to each other. Hashing 
organizes the data according to a mapping 
function which is used to hash each data item. Hashing is an efficient 
method to organize and search data 
quickly. The temporal efficiency or speed of the hashing process is 
determined by the hash function and its 
complexity. 

\vspace{3 mm}

Hash functions are mathematically based, and a common hash function uses 
the remainder mod  operation 
to map data items. Other hash functions are based on mathematical 
formulas and equations. The construction of 
a hash function is usually from multiplicative, divisional, additive 
operations, or some mix of them. The choice 
of hash function follows from the data items and involves temporal and 
space organization compromises.

\vspace{3 mm}

Hashing is not a direct mathematical mapping of the data items into a 
space organization. Hashing 
functions usually have a phenomena of a hash collision or clash, where 
two data items map to the same spatial 
location. This is detrimental to the hash function, and many techniques 
for handling hash collisions exist. Hash 
collisions introduce temporal inefficiency into a hash algorithm, as the 
handling of a collision represents 
additional time to re-map the data item.

\vspace{3 mm}

A special type of hash function, a perfect hash function, exists and is 
perfect in the sense it will not have 
collisions. These types of perfect hash functions are often available 
only for narrow applications, reserved 
words in a compiler symbol table for example. Perfect hash functions 
usually involve restricted data items and 
do not fit in to a general purpose hash method While perfect hashing is 
possible, often a regular hash function is 
used with data in which the number of collisions is few or the collision 
resolution method is simple.

\section{Algorithm}

\subsection{Super-Hash Function}

\indent\indent

The heart of the hash sort algorithm is the concept of combining hashing 
along with sorting. This key 
concept is embodied in the notion of a super-hash function. A super-hash 
function is "super" in that it is a hash 
function that has a consistent ordering, and is a perfect hash  function. 
A hash function of this type will order 
the data by hashing, but maintain a consistent ordering and not scatter 
the data. This hash function is perfect so 
that as the data is ordered it is uniquely placed when hashed, so that no 
data items ambiguously share a location 
in the ordered data. Along with that, the necessity for hash collision 
resolution can be avoided altogether.

\vspace{3 mm}

The super-hash function is a generalized function, in that it is 
extendible mathematically, and is not a 
specialized type of hash function. The super-hash function operates on a 
data set which is of positive integer 
values. With a super-hash function, the restrictions on the data set as 
the domain is that it is within a bounded 
range, between a minimum and maximum value. The only other restriction is 
that each integer value be unique 
-- no duplicate data values. This highly closed set of data values as 
positive integers is necessary to build a 
preliminary super-hash function. Once a mathematically  proven super-hash 
function is formulated, then other 
less restricted data sets can be explored.

\vspace{3 mm}

A super-hash function as described and within the parameters is not a 
complex or rigorous function to 
define. A super-hash function uses the standard hash function using the 
modulus or residue operator. This 
operator is integer remainder, called \textbf{mod}, is a common hash 
function in the form $\mathbf{(x\ mod\ n)}$. The other
part of 
the super-hash function is another hash function called a mash function, 
for modified hash or magnitude hash. 
The mash function uses the integer division operator. This operator, 
called \textbf{div}, is used in the form similar to the 
hash of  $\mathbf{(x\ div\ n)}$.  Both of these functions, the mash 
function and the hash function, together form the super-
hash function. This super-hash function is mathematically based and 
extensible, and is also a perfect hash 
function. For the super-hash function to be perfect, it must be an 
injective mapping.

\vspace{3 mm}

The super-hash function works using a combination of a regular hash 
function and the mash function. 
Together both of  these functions make a super-hash function, but not as 
the composition of the two. Both the 
hash function, and mash function are sub-functions of a  super-hash 
function.  The regular has function $\mathbf{(x\ mod\ n)}$ 
works using the remainder or residues of values. So numbers are of the 
form $c \cdot x + r$, where r is  the 
remainder obtained using the regular has function. When hashing by a 
value $n$, the resulting hashes map from 
the range 0 to $n - 1$. In essence, a set of values is formed so that 
each value in the set is $ \{ c\cdot x + 0, c \cdot x +
1, \ldots , c \cdot x 
+ (n - 2), c \cdot x + (n-1) \} $. 

\vspace{3 mm}

A hash function provides some distinction among the values, using the 
remainder or  residue of the value. 
However, regular hashing experiences collisions, or where values hash to 
the  same value. The problem is that 
values of the same remainder are indistinguishable to the regular hash 
function.  Given a particular remainder r, 
all  values which are multiples  of $c$ are equivalent. So a set of the 
form  $ \{ c_1 \cdot x + r_1,  c_2 \cdot x + r_1,
\ldots , c_{n-1} \cdot x+ r_1, 
c_n \cdot x + r_ 1 \}$. So for $n = 10$, $r = 1$ the following set of 
values are equivalent under the regular hash function: $
\{1, 11, 
21, 31, \ldots 1001, 10001, c \cdot x + 1 \}$. It is  the equivalence of 
the values under regular hashing which causes 
collisions, as values map to  the same hash result. There is no 
distinction  among larger and smaller values 
within the same hash result.  Some partitioning or separation of the 
values is obtained,  but what is needed is 
another hash function to distinguish the hash values further by their 
magnitude relative to one another in the 
same set.  

\vspace{3 mm}

This is what the mash function is for, a magnitude hash. A mash function 
is of the same form  as a regular 
hash function, only it uses div rather than mod for $\mathbf{(x\ mod\ 
n)}$. The \textbf{div} operator on the form of a value
$c \cdot x+r$ gives 
the value of  $c$, where $x$ is the base (usually decimal, base 10). The 
mash  function maps values into  a set where 
the values mash to the same result, based upon the magnitude. So the mash 
function shares the same problem  
as a regular hash that all values are mapped into an equivalent set.  So 
a set of the from $\{ c_1 \cdot x + r_1, c_1 \cdot x
+ r_2, \ldots , 
c_1 \cdot x + r_{n-1}, c_1 \cdot x + r_n \}$ has all values mashed to the 
same result. With n = 10, c = 3 the following set
of values 
are  equivalent  under the mash  function $\{ 30, 31, 32, 33, 34, 35, 36, 
37, 38, 39 \}$. With the mash function, 
some partitioning  of the values is obtained, but there is no  
distinction among  those values which is unique to 
the values. 

\vspace{3 mm}

Together, however, a hash function, and mash function can both 
distinguish values, by magnitude,  and by 
residue. The minimal form of this association between the two functions 
is an ordinal pair of the form $(c,r)$ 
where c is the multiple of the  base obtained with  the mash function,  
and r is the remainder of the value 
obtained with the hash function. Essentially an ordinal pair form a 
unique representation to the value, using the 
magnitude constant, and residue. Further, each pair distinguishes larger 
and smaller values, using the mash 
result, and equal magnitudes are distinguished from each other using the 
residue. So the mapping is a perfect 
hash, as all values are uniquely mapped to a result, and is ordering, 
since the magnitude of the values is 
preserved in the mapping. A  formal proof of this property, an injective 
mapping from one set to a resulting set, 
is given.

\vspace{3 mm}

The values for n in the hash function and mash function are determined by 
the range of values involved. 
The proof gives the validation of the injective nature of the mapping, 
and the mathematical properties of the 
parameters used, but no general guidelines for determining the 
parameters. The mathematical proof only 
indicates that the same mapping value must be used in the hash and mash 
functions. 

\vspace{3 mm}

Multiple iterations of  mapping functions can be used, so multiple 
mapping values can be used to form a set 
of mapping pairs. The choice of values for the mapping values depends on 
the number of dimensions to be 
mapped into, and the range of values that the mapping pairs can take on. 
Numerous, smaller mapping values 
will produce large mapping ordinates for the mash function, and small 
values for the hash function. This would 
be a diverse mix of values, but it depends upon the use of the hash sort 
and what is desired by the user of the 
algorithm. 

\vspace{3 mm}

The organization of the data elements in the matrix can be of row-major 
or column-major order. The mapping into a
column and row by the hash and mash function determines if the matrix is 
row or column major mapping. A row-major 
order mapping would have the rows mapped by the mash function, and the 
columns mapped by the hash function. A column
major order mapping would interchange the mash and hash functions for the 
rows and columns respectively.

\subsection{Construction of the Super-Hash Function}

\subsubsection{Method of Construction}

The super-hash function consists of a regular hash function using the 
\textbf{mod} operator,
and a mash function using the \textbf{div} operator. Together these form 
ordinal pairs 
which uniquely map the data element into a square matrix. The important 
component
of the super-hash function to determine is the mapping constant $\Theta$. 

\vspace{3 mm}

To determine $\Theta$, it must be calculated from the range of the values 
that are to be 
mapped using the super-hash function. The range determines the 
dimensionality or size
of the matrix, which is square for this super-hash function.

\vspace{3 mm}

Given a range $R [i,j]$ where $i$ is the lower-bound, and $j$ is the 
upper-bound, then:

\vspace{3 mm}

\begin{enumerate}

\item
Compute the length $L$ of the range where $L = (j - i) + 1$ .
\item
Determine the nearest square integer to $L$. The nearest square is 
calculated by:

   \[ \Theta = \lceil \sqrt{L} \rceil \]

\end{enumerate}

   The final value computed is the nearest square to $L$, which is the 
length of
the range of values. In essence the values are being constructed in the 
form of
a number which is tailored to the length of the range of values.

  \[ value_(d_x,m_x) = d_x \cdot \Theta + m_x \]

 where $\Theta$ is determined by the range of the values to be mapped.

\subsubsection{Example of Constructing a Super-Hash Function}

As an example, suppose you have a range of values from 13 to 123. The 
length of
the range is 123 - 13 + 1. The length of this range is 111. 

\vspace{3 mm}

The nearest square is then calculated as follows:

\[ \Theta = \lceil \sqrt{111} \rceil \]

which evaluates as:

\[ \Theta = \lceil  10.53565375 \ldots   \rceil \]

\[ \Theta = 11 \]

Part of the mapping involves subtracting the lower bound of the range, so 
that all 
the values are initially mapped to the range $0 .. (j - i)$. So the 
super-hash function
for this range of data values is:

\[
F(x) \equiv \left\{ \begin{array}{ll}
d & = (x - 13)\ \mathbf{div}\ 11\mbox{\ \ \ \ \ \ \ \ }\\
m & = (x - 13)\ \mathbf{mod}\ 11\mbox{\ \ \ \ \ \  }\\
\end{array} \right.
\]

\noindent\noindent
For the lowest value, 13 maps to (0,0). The largest value 123 maps to (10,0).

\vspace{3 mm}

Reconstructing the values from the ordinal pairs is of the form:

\[ value_(d_x,m_x) = d_x \cdot \Theta + m_x + i \]

so that for (0,0)

\[ value_(0,0) = ( 0 \cdot 11 + 0) + 13 = 13 \]

and for (10,0)

\[ value_(10,0) = ( 10 \cdot 11 + 0) + 13 = 123 \]

One final point is that constructing a super-hash function requires 
knowledge of
the range of possible values of the data elements. All data types in 
computer languages
have a specific range of values for the different types, such as int, 
float, unsigned int,
long to list types from the C programming language. While these types are 
often used without
needing to know the range of values, there does exist such a property on 
those values and 
variables defined of that particular type in a program.

\subsection{In-situ Hash Sort Algorithm}

\indent

The hash sort algorithm uses the super-hash function iteratively on an 
entire data set within  the range of the 
super-hash function. This is the in-situ version of the hash sort, which 
works "in site". 
Before the iterative process, an initialization is performed.  A source 
value is retrieved, and 
is mapped by super-hash  to another location. At the destination 
location, the destination  value is  exchanged 
with the source one, and stored at the location. The new value now is a 
source value. The algorithm process 
then repeats iteratively for a source, destination  value. The algorithm 
terminates at the end of the list.

\vspace{ 3 mm}

\noindent
Pseudo-code illustrating a very generalized from of the in-situ hash sort is:

\[ ( m_1, m_2, \cdots ,m_{n-1},m_n ) \longleftarrow initialize; \ \ \ \ \ 
\ \ \ \ \ \ \ \ \ \  \] 
\[ \textbf{ WHILE NOT } (\ end\_of\_list \ ) \textbf{ DO } \ \ \ \ \ \ \ 
\ \ \ \ \ \ \ \ \ \ \ \]
\[ temp \longleftarrow get( m_1, m_2, \cdots ,m_{n-1}, m_n );\ \]
\[ \ \ value \longrightarrow put( m_1, m_2, \cdots ,m_{n-1}, m_n);\ \ \]
\[ \ \ \ value = temp\ ; \ \ \ \ \ \ \ \ \ \ \ \ \ \ \ \ \ \ \ \ \ \ \ \ 
\ \ \ \ \ \ \ \ \  \]
\[ \ \ \ \ \ \ \ \ \ \ \ \ \ \ (\ m_1, m_2, \cdots ,m_{n-1}, m_n \ ) 
\longrightarrow super\-hash(temp); \]
\[ \textbf{  END WHILE ;\ \ \ \ \ \ \ \ \ \ \ \ \ \ \ \ \ \ \ \ \ \ \ \ \ 
\ \ \ \ \ \ \ \ \ \ \ \ \ \ } \]

\vspace{3 mm}

\noindent\noindent
Pascal Version:

\vspace{3 mm}

\begin{verbatim}
Procedure hash_sort(var list: data_list; n: integer); 
Const
  D=10;  (* D is the dimension size for a 10 x 10 matrix *)

Var
  x_c, h_c: integer;      
  which, where: integer;  
  value: integer;         
Begin
  x_c := 0;   (* set the counts to zero *)
  h_c := 0;
  where := 0; (* set the initial starting points to zero *)

  while( (x_c < n) AND (h_c < n) ) do
  begin    (* loop until exchange and hysteresis count equal data size *)
     value := List[where div i, where mod j];  (* get a value *)
     if (value = where) then  (* check for hysteresis *)
     begin
       where := where + 1;  (* on hysteresis move where to next position *)
       h_c := h_c + 1;      (* on hysteresis increment hysteresis count *)
     end
     else   (* if no hysteresis, swap values and increment exchange count *)
     begin
        List[where div D, where mod D] := List[value div D, value mod D];
        List[value div D, value mod D] := value;
        x_c := x_c + 1;
     end;
   end;
End;
\end{verbatim}

\vspace{3 mm}

\noindent
C version:

\vspace{3 mm}

\begin{verbatim}
#define DIM 10 /* dimension size for the matrix */

void hash_sort(int& list[DIM][DIM], int n){
   int x_c = 0,h_c = 0;
   int which = 0,where = 0;
   int value;

   while( (x_c < n) && (h_c < n) ){
      value = list[where / DIM][where % DIM];
      if (value == where){
        where++;      /* on hysteresis move where to next position */
        h_c++;        /* on hysteresis increment hysteresis count */
      } else {
        /* if no hysteresis, swap values and increment exchange count */
        list[where / DIM][where % DIM] = list[value / DIM][value % DIM];
        list[value / DIM][value % DIM] = value;
        x_c++;        /* increment the exchange count */
      }
   }
}
\end{verbatim}

\noindent
Generic Pseudo-code of operations involved in in-situ version of algorithm:

\begin{verbatim}
PROCEDURE HASH_SORT;
    Initialize variables;

    WHILE ( Counts < Data Size ) DO
        Mathematically process value;
        Compute destination in array;

         IF hysteresis THEN
              Move where to next location;
              Increment hysteresis count;
         ELSE
              Exchange value with destination;
              Increment exchange count;
         END IF;
    END WHILE;

END PROCEDURE:
\end{verbatim}

\vspace{3 mm}

The hash sort is a  simple algorithm, the complexity  is in the 
super-hash function which is the key to the 
algorithm. The remaining components of  the algorithm are for  the 
exchange of values, and continuous 
iteration until a termination condition.

\vspace{3 mm}

The hash sort is a linear time complexity sorting algorithm. This linear 
time complexity stems from the 
mapping nature of the algorithm. The hash sort uses the intrinsic nature 
of  the values to map them in order. 
The mapping function is applied repeatedly in an iterative manner until  
the entire list of  values is ordered. The 
mathematical expression for  the run-time complexity is:

\vspace{3 mm}
\[ F(time) = 2\cdot c\cdot N \]
\noindent
where $c \geq 1$ , and $N$ is the size of the list of data values.

\vspace{3 mm}

The function for the run-time complexity is derived from the complexity 
of the mapping  function, and the 
iterative nature of  the algorithm. The mapping function  is a composite 
function, with two sub-functions. 
Multiple applications of the mapping function are possible because of the 
extendible nature of the hash sort, but 
at least one mapping is required. Hence, the  constant $c$ is a positive 
integer value, which represents  the number 
of sub-mapping within the mapping function.

\vspace{3 mm}

The mapping function uses two sub-functions as a composite, so the 
overall time for the mapping is the 
product of two multiplied by the number of  sub-mappings. The value of 
the product of two multiplied by the 
number of  sub-mappings is always greater than one,  hence it is the 
dimension  of the hash sort. The constant c 
is not  dependent on the data values or the size of the data values, it 
is an implementation constant; once  the  
constant  is chosen it is unaltered.

\vspace{3 mm}

The mapping  function must be applied iteratively to the range of values 
forming the data list. This makes 
the overall time complexity of the hash sort the product of the 
complexity of the mapping function multiplied 
by the size of the range of the list of values. The hash sort remains 
multiple dimension, and linear in time 
complexity. The time complexity of the hash  sort is then:

\[ F(time) = O(N) \]

\vspace{3 mm}

The space complexity of the hash sort is considering  the storage of  the 
data values, and  not the variables 
to used in the mapping process. The space complexity of the hash sort is 
dependent upon which variant of the 
hash sort is used. The in-situ hash sort,  which uses the same 
multi-dimensional data structure as it maps the 
values from the initial location to the final location, requires $(N + 
1)$ storage space. The data  structure is the 
size of the range of  values, which is $N$. One additional storage space 
is required as a temporary storage location 
as values are exchanged  when mapped.  The space  complexity for the 
in-situ hash sort is:

\[ F(space) = O(N) \]

\vspace{3 mm}

The direct hash sort maps from a one-dimensional data structure to the 
multi-dimensional data structure of 
the same size. No temporary storage is required, as no values are 
exchanged in the mapping  process,  the 
values are directly mapped to the final location within the  
multi-dimensional data structure. The two data  
structures are organized differently, but are of the same size $N$. 
Thus,  the direct hash sort requires $2 \cdot N$ in terms 
of space complexity. The space complexity for the direct hash sort is:

\[ F(space) = O(N) \]

\vspace{3 mm}

Each variation of  the hash  sort has different properties, but the space 
complexity is $(N + 1)$ for the in-situ 
hash sort, and $2\cdot N$ for  the direct hash sort. For  each variation, 
the space complexity is linearly related to the 
size of the  list of data values.

\vspace{3 mm}

The hash sort performance asymptotically outperforms conventional sorting 
algorithms (such as quick sort), which are $N log N$
time complexity
performance. This is readily apparent from  a simple inequality involving 
the ratio of the two algorithms. As the size of 
the data $N$ increases without bound, then the ratio between the hash 
sort and a quick sort should be less than one.

\vspace{3 mm}

If the ratio is greater than one, the time complexity of the hash sort is 
greater than the quick sort. If the ratio is exactly
one then the two sort algorithms perform with the same complexity. A 
ratio of less than one indicates the hash sort is less than
the time complexity of the quick sort, therefore outperforming it.

\[ 
\lim_{N \to +\infty}
\frac{2 \cdot k \cdot N + c}{N \cdot log N}\ < 1.0
\]

which simplifies to:

\[
\lim_{N \to +\infty}
\frac{2 \cdot k  + \frac{c}{N}\ }{ log N }\ < 1.0
\]

taking the $\lim_{N \to +\infty}$ the ratio then becomes:

\[
\frac{2 \cdot k  + \frac{c}{\infty}\ }{ log \ \infty }\ < 1.0
\]

which simplifies to:

\[
\frac{2 \cdot k }{ \infty }\ < 1.0
\]
	
which then reduces to:
\[
0.0 < 1.0
\]

\noindent\noindent

showing that the ratio of the two algorithms is indeed less than one. 

\vspace{3 mm}

Therefore this means the hash sort will asymptotically outperform the 
quick sort.

\subsection{Variations of the Hash Sort}

\indent\indent
There are two types of versions of the hash sort, the in-situ hash sort, 
and the direct hash  sort. Variations 
upon the hash  sort are upon these two primary types. The  in-situ hash 
sort is the basic form of the hash sort, 
which works by exchanging values as explained previously. The in-situ 
hash sort has a problem which can 
increase its time-complexity,  but the hash sort algorithm  remains 
linear. The in-situ hash sort has a problem 
with data values that map  to their current location.  In essence, the 
data value is already in-site, where it 
belongs. Since the in-situ hash sort will determine where it belongs, 
then exchange, this would cause the in-situ 
hash sort to halt. To remedy this, another iterative mechanism keeps the 
in-situ hash sort algorithm going by 
relocating the current location to the next one. When the current 
location and destination  location are the same, 
the in-situ hash sort has to  be forced to  proceed.

\vspace{3 mm}

This forcing of the in-situ hash  sort does add more time-complexity, but 
as a linear multiplied constant. 
The number of data elements that map back to the current location  they 
are at is the amount of  hysteresis 
present in the data. The term hysteresis is borrowed from electrical 
theory, meaning to lag; hysteresis in the  in-
situ  hash sort causes a lag, which the algorithm must be forced out of. 
The worst case for hysteresis is that all 
of the data elements map to the location they are  at. In this case, the 
in-situ hash sort would have  to be pushed 
along until it  is through the data elements. In this worst case, the 
time of the algorithm becomes a double of the  
linear time complexity, increasing the time, but linearly to $2(2 \cdot 
k) \cdot N$, or  $4 \cdot k \cdot N$. The  in-situ
hash sort is more 
space efficient, using only the space to store the data elements, and a 
temporary storage for sorting the data 
elements,  or $N + 1$. 

\vspace{3 mm}

The direct hash sort  is a variation upon the  in-situ to avoid the 
problem with hysteresis. The  direct hash 
sort uses a temporary data array to hold the data elements. The  data 
elements are then mapped from the single 
dimension array into the  multiple dimension  array. No element can map 
to its current location, as it is being 
mapped from a  one-dimensional array into a multiple dimensional. 
However, the storage requirements for the 
direct hash sort require twice the storage, or $2 \cdot N$. The tradeoff 
for time efficiency is a worsening of the space 
efficiency.  The time complexity is $2 \cdot k \cdot N$, as the  problem  
with hysteresis never surfaces. 

\vspace{3 mm}

The variations in the hash sort can be applied to the two primary forms, 
the in-situ and the direct  hash sort. 
The variations are in dimensionality, and relaxing a restriction imposed 
upon the data set that is sorted. The 
version of the hash sort mentioned is  two-dimensional, but the hash sort 
can be of any dimension $d$, where $d \geq 2$. 

\vspace{3 mm}

The map by hashing would then be multiple  applications  of the hash 
scheme. For a d-dimensional hash 
sort, it will be of the form $2 \cdot k \cdot N$, where $k$ is the 
dimensionality of the  hashing structure. Note that $k$ is
an 
implementation constant, so once decided upon, remains a constant 
multiple of 2. Either primary type of hash 
sort algorithm, the in-situ hash sort, or the direct hash sort,  can be 
extended into higher dimensionality.

\vspace{3 mm}

The other variation upon the  primary forms of the  hash sort  concerns 
the restriction of unique data values. 
This restriction upon the data set can be relaxed, but with  each 
location in the hash structure a count is required 
of the number of data elements that hash to that location. For large 
amounts of values in the data set, the  hash 
sort will "compress" them into the hash structure. When the  original 
sorted data set is required, each data 
element would be enumerated by the number of hashes to its location. The 
flexibility of the hash sort to 
accommodate non-unique data values is inefficient if there are few 
repeated data values. In such  a case,  
nearly half of the hash structure will be used only for single data 
values. 

\vspace{3 mm}

Both the in-situ hash  sort and the direct hash sort have another 
problem, which  is  inherent in either variant 
of the  hash sort algorithm. If the data values within the  range are not 
all there, then the data set is sparse. The 
range is used to determine the hash structure size, and the time to run 
the algorithm. The hash sort presumes all 
data values  in the data  set within the range are present. If not, then 
the hash  sort  will  still proceed sorting for  
the range size, not the data set size. So for the data set size $N_d$, 
and the  range  size $N_r$, if $N_d \geq N_r$, the hash
sort 
algorithm performs as expected, or better. 

\vspace{3 mm}

If  $N_d < N_r$, then the sparsity problem surfaces. The  hash sort will 
sort on a range of values, of which some are
non-existent. The smaller data set size to range size will not be 
reflected in the time-complexity of the hash sort,  or in the required 
hash structure; so when the cardinality of 
the data set and cardinality of  the  range of values  within the data 
set are inconsistent, the hash sort doesn't 
flop, but it becomes inefficient. The empty spaces within the  hash 
structure then must have some sentinel value 
outside the data range  to distinguish them from  actual data values. The 
direct hash sort alleviates the sparsity 
problem somewhat for the hash sort time,  but still is inefficient in the 
hash structure as some of the locations 
will be blank or empty. 

\vspace{3 mm}

\noindent\noindent
Code for the direct hash sort algorithm is:

\vspace{3 mm}

\noindent\noindent
C version:

\begin{verbatim}

#define DIM 10  /* dimension size of the matrix */

typedef struct tag {
   int count;
   int value;
} element;

element matrix M[DIM][DIM]; 

void hash_sort(list L[], element M[ ][DIM], int size)
{
     int value,x;
     int row, col;

     for(x=0;x < size;x++)
     {
        /* get the value and detemine its row, column location */
        value = L[x];
        row = value % DIM;
        col = value / DIM;  

        if(M[col][row] == value)
          /* if the value is already here, increment count */
          M[col][row].count++;
        else {
          /* store the value, initialize the count to 1 */
          M[col][row].count = 1;
          M[col][row].value = value;
        }
     }
}

\end{verbatim}     

\noindent\noindent
Pascal version:

\begin{verbatim}

const
    d = 10;
    size = 100;
type
   record tag =
     count: integer;
     value: integer;
   end;

   matrix = array[d,d] of tag;
   list = array[size] of integer;

procedure hash_sort(var matrix M; list L; integer size)
var
    value: integer;
    row, col: integer;
    x: integer;
begin
    (* go through list and map values into matrix *)
    for x:= 1 to size do begin
        (* get current value, determine row and column location *)
        value:=L[x];
        row:=value div x;
        col:=value mod x;

        if M[col,row]=value then
            (* if value already here *)
            M[col,row].count:=M[col,row].count + 1;
        else begin
            (* if maps first time, store value, initialize count to 1 *)
            M[col,row].count:=1;
            M[col,row].value:=value;
        end; (*if*)

    end; (*for*)

end; (*hash_sort*)


\end{verbatim}

\subsection{Differences between Hash Sort Variants}

\subsubsection{Distinction of Concepts}

There are three distinct concepts involved in the hash sort which need to be
distinguished from one another. Each has an important significance 
relating 
to the hash sort, and is explained as it is identified.

\vspace{ 3mm}

These three concepts are:

\vspace{3 mm}

\begin{enumerate}

\item
Number of data elements $N$ 
\item
Range of the data values from a lower to upper bound $R$
\item
Square matrix which is the data structure $M$

\end{enumerate}

\vspace{3 mm}

The number of data elements $N$ is the total data size to be mapped by the
hash sort. The number of data elements determines the time-complexity of the
hash sort, the amount of time being linearly proportional to the number 
of 
data elements. 

\vspace{3 mm}

The in-situ hash sort, the number of data elements is less or equal to 
the size 
of the square matrix $M$. The direct hash sort, the number of elements 
can be of 
any size. The only requirement is that the data elements fall within the 
range $R$. 

\vspace{3 mm}

The square matrix $M$ is constructed around the range $R$ of the data 
values so that
all possible values within the range $R$ do map. The super-hash function 
maps data 
elements within the range $R$ into the square matrix $M$. Therefore, the 
storage space
$M$ is formed from the range $R$ of the data values, not the number of 
data elements $N$.

\vspace{3 mm}

The time complexity is dependent upon the number of data elements, the 
size $N$.
The in-situ hash sort can have less elements $N$ than the matrix capacity 
$M$, but the
hash sort algorithm must go through the entire matrix $M$. Hence sparsity 
of data values
within the matrix is highly inefficient. The direct hash sort is 
different since the
mapping is from a list into the matrix $M$. The time complexity is linear 
again to the
size $N$ of the list of elements. 

\vspace{3 mm}

In both cases, there is linear time complexity, only with the in-situ, it 
is dependent 
more upon the size of the matrix $M$ than the amount of data in the 
matrix. In doing so, 
the time complexity is more dependent upon the range of values for the 
in-situ than the 
number of them. No more data elements can be stored in the matrix than 
its size $M$ permits, 
which is related to the range $R$.

\subsubsection{Analysis of Hash Sort Variants}

Depending upon the variant used, the hash sort can have a time complexity 
linearly proportional 
to either the size of the data structure which is a matrix $M$, or the 
size of the data list $L$. In
each variant of the hash sort, the linear nature of the algorithm is 
proportionate to a greatest
lower bound. 

\vspace{ 3mm}

The in-situ variant of the hash sort is linearly proportional to the size 
of the
data structure, the matrix $M$. The matrix $M$ has a size determined in 
part bythe super-hash
function. The size of the matrix places a least upper bound on the 
possible size of the list $L$,
and in doing so forms a least upper bound of $O(M)$ time complexity. This 
least upper bound 
constraint stems from the fact that the data list $L$ and the data 
structure the matrix $M$ are
the same entity.

\vspace{ 3mm}

The direct variant of the hash sort is linearly proportional to the size 
of the data list $L$, as 
the data structure the matrix $M$ and the list $L$ are seperate entities. 
The time complexity is
independent of the data structure in the direct hash sort. This 
independence of entities permits
a greatest lower bound dependent on the size of the data list $L$ and not 
the data structure the
matrix $M$.

\vspace{ 3mm}

For each variant of the hash sort, the time complexity of the algorithm 
is linear. The linearity is 
constrained in relation to two seperate determining factors for each 
variant of the hash sort. 
In one variant, the in-situ version of the hash sort, the time complexity 
has a least upper bound determined by the
matrix $M$. The size of the data list $L$ in $M$ can be less than, but no 
more than the size of 
the matrix $M$ because the data list $L$ and the data structure $M$ are 
the same entity. The direct
hash sort variant seperates these two entities and in doing so the 
constraint is a greatest lower 
bound dependent upon the size of the data list $L$. 

\vspace{ 3mm}

A table summarizing these distinctions is given below:

\vspace{ 3mm}

%table of time complexity of features

\begin{tabular}{|c|c|c|c|}
\hline
Variant & Big-Oh & Constraint & Comment\\
\hline\hline
In-situ & O(M) & Data Structure Size & $O(M) \le O(N)$ \\ 
\hline
Direct & O(N) & Data List Size & $0 \le O(N)$ \\
\hline
\end{tabular}

\vspace{ 3mm}

\subsubsection{Example Walk-through of Hash Sort Variants}

\begin{verbatim}

 In-situ hash sort example with 2-dimensional 3 x 3 matrix



                              Matrix:

                            m = 0  1  2
                          d
                           0    *  *  *
                           1    *  *  *
                           2    *  *  *


\end{verbatim}

The values in the matrix must be mapped to the range 0 .. 8; this range 
of values forms a ordinal 
pair of the form $(d,m)$ from (0,0) to (2,2). Values in the matrix 
increase from left to right 
m = 0 .. 2, and top to bottom d = 0 .. 2. This is how the ordering should 
place values once they are 
sorted.

\vspace{3 mm}

Any range of values can be used, but for simplicity, the range of values 
in the matrix is from 1
to 9. The hash sort algorithm used will be in-situ (or in site) so the 
problem of hysteresis is
present. Hysteresis is the occurrence where a value is in its correct 
location, so the algorithm
maps it back on to itself. 

\vspace{3 mm}

The hash sort will involve starting at some initial point in the matrix, 
then traversing the 
matrix and mapping each value to its correct location. As each values is 
mapped, order is 
preserved (a large value will map to a "higher" position than a "lower" 
one), and there are no 
collisions (excepting self-collisions which are hysteresis). The values 
are mapped by subtracting 
one, then applying the \textbf{mod} and \textbf{div} operators.

\vspace{3 mm}

The where, or which digit is being handled is in parentheses, the 
computed destination is in 
brackets. The value used in the computation is given, and the computed 
ordinal pair. A before and 
after illustration is given to show the exchange of the two values which 
is computed, then done 
by the hash sort.

\vspace{3 mm}

\noindent\noindent
The super-hash function for this data set is:

\[ d \ = (x-1) \ \textbf{div} \  3\ ; \  m \ = (x-1) \  \textbf{mod} \ 3 \]

\begin{verbatim}


                             m = 0  1  2
                          d 
                            0    5  8  1
                            1    9  7  2
                            2    4  6  3


                     Matrix Initial Configuration



            m = 0  1  2                       m = 0  1  2				
         d                                 d
           0   (5) 8  1                      0   (7) 8  1
           1    9 [7] 2                      1    9  5  2
           2    4  6  3                      2    4  6  3

                Before                            After


\end{verbatim}

\noindent\noindent
The start position is initially at (0,0). The value is 5, subtracting 1 
is 4. 
The mapping (illustrated once) is d = (4 div 3) = 1, m = (4 mod 3) = 1.
Thus the computed destination for where the value goes is (1,1). 

\begin{verbatim}
	


            m = 0  1  2                       m = 0  1  2			
         d                                 d
           0   (7) 8  1                      0   (4) 8  1
           1    9  5  2                      1    9  5  2
           2   [4] 6  3                      2    7  6  3

               Before                             After

\end{verbatim}

\noindent\noindent
The next value is 7, subtracting 1 is 6. The computed destination for 
where the value goes is 
(2,0). 

\begin{verbatim}
	
            m = 0  1  2                       m = 0  1  2				
         d                                 d
           0   (4) 8  1                      0   (9) 8  1
           1   [9] 5  2                      1    4  5  2
           2    7  6  3                      2    7  6  3

               Before                             After

\end{verbatim}

\noindent\noindent
The next value is 4, subtracting 1 is 3. The computed destination for 
where the value goes is 
(1,0). 

\begin{verbatim}
	
            m = 0  1  2                       m = 0  1  2				
         d                                 d
           0   (9) 8  1                      0   (3) 8  1
           1    4  5  2                      1    4  5  2
           2    7  6 [3]                     2    7  6  9

                Before                            After

\end{verbatim}

\noindent\noindent
The next value is 9, subtracting 1 is 8. The computed destination for 
where the value goes is 
(2,2). 

\begin{verbatim}
	
            m = 0  1  2                       m = 0  1  2				
         d                                 d
           0   (3) 8 [1]                     0   (1) 8  3
           1    4  5  2                      1    4  5  2
           2    7  6  9                      2    7  6  9

                Before                            After

\end{verbatim}

\noindent\noindent
The next value is 3, subtracting 1 is 2. The computed destination for 
where the value goes is 
(0,2). 

\begin{verbatim}
	
            m = 0  1  2                       m = 0  1  2				
         d                                 d
           0   <1> 8  3                      0    1 (8) 3
           1    4  5  2                      1    4  5  2
           2    7  6  9                      2    7  6  9

                Before                            After

\end{verbatim}

\noindent\noindent
The next value is 1, subtracting 1 is 0. The computed destination for 
where the value goes is 
(0,0). 

\vspace{3 mm}

\noindent\noindent
Here is an example of hysteresis noted by the angular brackets; the start 
and final positions are 
equal, so the value has mapped back onto itself. The start position is 
'forced' to the next 
position or the algorithm would be stuck in an infinite loop and so would 
"stall".

\begin{verbatim}
	
            m = 0  1  2                       m = 0  1  2				
         d                                 d
           0    1 (8) 3                      0    1 (6) 3
           1    4  5  2                      1    4  5  2
           2    7 [6] 9                      2    7  8  9

                Before                            After

\end{verbatim}

\noindent\noindent
The next value is 8, subtracting 1 is 7. The computed destination for 
where the value goes is 
(2,1). 

\begin{verbatim}

            m = 0  1  2                       m = 0  1  2				
         d                                 d  
           0    1 (6) 3                      0    1 (2) 3
           1    4  5 [2]                     1    4  5  6
           2    7  8  9                      2    7  8  9

                Before                            After

\end{verbatim}

\noindent\noindent
The next value is 6, subtracting 1 is 5. The computed destination for 
where the value goes is 
(1,2). 

\begin{verbatim}

            m = 0  1  2                       m = 0  1  2				
         d                                 d
           0    1 <2> 3                      0    1  2  3
           1    4  5  2                      1    4  5  6
           2    7  8  9                      2    7  8  9

                Before                            After

\end{verbatim}

\noindent\noindent
The next value is 2, subtracting 1 is 1. The computed destination for 
where the value goes is 
(0,1). 

\vspace{3 mm}

\noindent\noindent
Here is another example of hysteresis; the start and final positions are 
equal, so the value has 
mapped back onto itself. The start position is 'forced' to the next 
position or the algorithm 
would remain stuck.

\vspace{3 mm}

The matrix is now sorted, but the algorithm has no way of knowing this. 
It will continue from 3 
to 9 until the hysteresis count equals the size of the data $N$. The 
remaining sorting data has 
caused "hysteresis" by causing the algorithm to lag in a sense, getting 
tripped up over already 
sorted data. In an ideal arrangement of the data, no hysteresis would 
occur; then the algorithm 
knows it has finished sorting with the exchange count equals the size of 
the data $N$. But 
unfortunately, this is an ideal situation which most likely will never 
occur in practice.

\begin{verbatim}

        Direct Hash Sort Example with 2 x 2 Matrix

                            Matrix

                        m = 0     1
                    d 
                      0   <*,*> <*,*>

                      1   <*,*> <*,*>

\end{verbatim}

The values of the data set are from 7 to 10. A list $L$ of size 7 of data 
elements to 
be mapped will be used. Each location in the matrix is the value, and an 
associated
count of the number of values in that location. Once all the values from 
the list $L$
are mapped, the hash sort algorithm will terminate.

\vspace{3 mm}

The hash function is:

\[ d \ = (x - 7) \ \textbf{div} \  2 \ ; \  m \ = (x - 7) \  \textbf{mod} 
\ 2 \]

As each value is mapped, an brief explanation of the process and what 
happens to the
matrix will be given. As the list L is mapped, it will become 
progressively smaller 
in representation through the walk-through. Once the list size is zero, 
the hash sort
will be complete. The left-most element in the list L will be the one 
being mapped by
the hash sort algorithm.

\begin{verbatim}

               L = { 7, 8, 7, 9, 10, 7, 8, 8 } 


                            Matrix

                        m = 0     1
                    d 
                      0   <*,*> <*,*>

                      1   <*,*> <*,*>

                 Matrix Initial Configuration
                            Step 0

\end{verbatim}

The first value is 7. The hash sort maps the value to the location in the 
matrix as 
d = (7 -7) div 2, m = (7-7) mod 2, which is (0,0)

\begin{verbatim}

               L = { 8, 7, 9, 10, 7, 8, 8 } 


                            Matrix

                        m = 0     1
                    d 
                      0   <7,1> <*,*>

                      1   <*,*> <*,*>

                            Step 1

\end{verbatim}

The next value is 8. The hash sort maps the value to the location in the 
matrix as
d = (8-7) div 2, m = (8-7) mod 2, which is (0,1).

\begin{verbatim}

               L = { 7, 9, 10, 7, 8, 8 } 


                            Matrix

                        m = 0     1
                    d 
                      0   <7,1> <8,1>

                      1   <*,*> <*,*>

                            Step 2

\end{verbatim}

The next value is 7. The hash sort maps the value to the location in the 
matrix as
d = (7-7) div 2, m = (7-7) mod 2, which is (0,0).

\begin{verbatim}

               L = { 9, 10, 7, 8, 8 } 


                            Matrix

                        m = 0     1
                    d 
                      0   <7,2> <8,1>

                      1   <*,*> <*,*>

                            Step 3

\end{verbatim}

The next value is 9. The hash sort maps the value to the location in the 
matrix as
d = (9-7) div 2, m = (9-7) mod 2, which is (1,0) 

\begin{verbatim}

               L = { 10, 7, 8, 8 } 

                            Matrix

                        m = 0     1
                    d 
                      0   <7,2> <8,1>

                      1   <9,1> <*,*>

                            Step 4

\end{verbatim}

The next value is 10. The hash sort maps the value to the location in the 
matrix as
d = (10-7) div 2, m = (10-7) mod 2, which is (1,1).

\begin{verbatim}




              L = { 7, 8, 8 } 

                            Matrix

                        m = 0     1
                    d 
                      0   <7,2> <8,1>

                      1   <9,1> <10,1>

                            Step 5

\end{verbatim}

The next value is 7. The hash sort maps the value to the location in the 
matrix as
d = (7-7) div 2, m = (7-7) mod 2, which is (0,0).

\begin{verbatim}



              L = { 8, 8 } 

                            Matrix

                        m = 0     1
                    d 
                      0   <7,3> <8,1>

                      1   <9,1> <10,1>

                            Step 6

\end{verbatim}

The next value is 8. The hash sort maps the value to the location in the 
matrix as
d = (8-7) div 2, m = (8-7) div 2, which is (0,1).

\begin{verbatim}



              L = { 8 } 

                            Matrix

                        m = 0     1
                    d 
                      0   <7,3> <8,1>

                      1   <9,1> <10,1>

                            Step 7

\end{verbatim}

The next value is 8. The hash sort maps the value to the location in the 
matrix as
d = (8-7) div 2, m = (8-7) div 2, which is (0,1).

\begin{verbatim}

              L = { } 

                            Matrix

                        m = 0     1
                    d 
                      0   <7,3> <8,2>

                      1   <9,1> <10,1>

                 Matrix Final Configuration

\end{verbatim}

All 7 data elements have been mapped by the hash sort into the matrix. 
There is no 
hysteresis with the direct hash sort, as all data elements are mapped by 
value into
the appropriate location. Moreover, the time of the direct hash sort is 
linearly 
proportional to the size of the data list $L$. 

\subsection{Other Similar Algorithms}

\indent\indent

There are three other algorithms that  have very strong similarities to 
the hash sort algorithm. These 
algorithms are: address calculation sort, bin sort, and the radix sort. 
While similar, these algorithms are distinct 
from the hash sort. 

\subsubsection{Address Calculation Sort:}

\indent\indent

The address calculation sort is very similar to the  hash sort. The 
address calculation sort  is sometimes 
referred to as sorting by hashing. The address calculation sort uses a 
hashing method that is order-preserving, 
similar to the  hash sort. However, the address calculation has the 
problem that if the distribution is not  
uniformly distributed, then the address calculation sort degenerates 
into  an $O(N^2)$ time complexity. This is the  
sparsity problem with the hash sort,  but it does not lead to such an 
extreme degeneration in the hash sort as it 
does with the address calculation sort. Another variant of the address 
calculation sort is the pigeonhole sort, in 
which the data list of elements is subdivided into bins, and then within 
each bin the sub data list is sorted.

\subsubsection{Bin Sort:}

\indent\indent

Bin sort is similar to hash sort in that data is stored in a "bin" which 
it is mapped to.  The bin sort, has 
multiple distinct values mapping to a similar bin. Unlike the hash sort, 
where redundant data values map to the 
same  location, the bin sort has distinct elements possibly mapping to 
the same bin. So the bin sort within each 
bin has multiple data  elements within the same bin. If there a N 
elements, and  M bins, then the bin sort  is  
linear time $O(N + M$). However, if the number of bins is $N^2$ , then 
the bin sort will degenerate into a worst case of 
$O(N^2)$. 

\subsubsection{Radix Sort:}

\indent\indent

The radix sort is similar to the hash sort in that the digits or 
sub-elements of each data value are used in the 
sort. The algorithm uses the digits of the data element to map it to its 
unique location. Hash sort does this 
indirectly not by each sub-element,  but by mathematical mapping. Radix 
sort for m-sized data element with  n 
elements has a time-complexity of $O(M \cdot N)$. If the sub-data 
elements become very dense, then m becomes more 
approximately $log N$, then the radix sort degenerates to a $O(N \cdot 
log\  N)$ algorithm. So hence, the radix sort depends 
on $M$ much less than $N$ by a sizable ratio.  

\subsubsection{Summary of Similar Algorithms}

\indent\indent

The similarity between the address calculation sort, bin sort, and radix 
sort is that a non-comparative 
method for sorting is used. However, all three  algorithms  degenerate 
into worst case in the very least being a 
comparative sort algorithm, or polynomial time algorithm.  This worst 
case  occurs at the extremes of  data 
density,  either too sparse or too dense, which then overloads the 
algorithm. The hash sort, while not incapable 
of degenerating, only becomes worst case of another linear time constant. 
So the hash sort is not as sensitive to 
extreme cases, and is more robust.

\subsection{Features of Hash Sort}

The hash sort has the following strengths:

\begin{itemize}

\vspace{3 mm}

\item Linear time complexity, even in the worst case; for the in-situ 
proportional to the data structure size,
and for the direct proportional to the data list size.

\vspace{3 mm}

\item The hash sort puts data elements in the correct position, does not 
move them afterward -- data quiesence

\vspace{3 mm}

\item Data independence -- data elements in the data set are mapped by 
their unique value, and do not depend on 
the predecessor or successor in the data set

\vspace{3 mm}

\item High speed lookup is possible once the data is sorted  -- faster 
than binary search; or alternatively, to
the approximate location within the data structure.

\vspace{3 mm}

\end{itemize}

\noindent
The hash sort has the following weaknesses:

\begin{itemize}

\vspace{3 mm}

\item Sparsity of data values in range -- wasteful of space

\vspace{3 mm}

\item Multi-dimensional data structure is required -- square planar 
matrices are inconsistent with 
underlying linear memory in one-dimension

\vspace{3 mm}

\item Works only with numeric values, requires conversion for non-numeric 
value

\vspace{3 mm}

\item The data range of values must be known for the algorithm to work 
effectively

\vspace{3 mm}

\end{itemize}

\section{Testing}

\subsection{Testing Methodology}

\indent\indent

The testing methodology of the sort algorithm involves two perspectives, 
an empirical or quantitative 
viewpoint, and a mathematical or qualitative view. The mathematical or 
qualitative point of view looks at the 
hash sort algorithm alone. The algorithm is tested for its characteristic 
behavior; this testing can be exhaustive, 
as there are an infinite number of sizes of test cases to test the 
algorithm with. The mathematical methodology 
then, tests for data size, and data arrangement or structuring. These 
form tests for which the size of the test data 
list is increasing, and tests in which there are partially sorted 
sublists within the overall test data list.
For simplicity, the testing emphasis was placed on the in-situ version of 
the hash sort.

\vspace{3 mm}

The testing approach for determing the behavior of the hash sort 
algorithm focus on the algorithmic time 
complexity. Testing is non-exhaustive, as all possible test cases can not 
be generated or tested upon. The 
algorithmic behavior is tested on different test cases to avoid the 
possibility of an anomalous "best" case or 
extreme "worst" case scenario. The hash sort algorithm is tested on 
different sizes and different permutations of 
data lists to evaluate the algorithmic performance. Complete data lists 
of unsorted, or fully sorted lists are used, 
along with partially sorted data lists. The hash sort is also compared to 
other algorithms that sort, to give 
relative comparisons and contrasts for better assessment. 

\vspace{3 mm}	

The two other sorting algorithms used to compare and contrast the hash 
sort are the bubble sort, and the 
quick sort. Each algorithm represents an extreme in algorithmic 
performance. The bubble sort is an O(n2) 
algorithm, but has excellent, linear $O(N)$ performance on partially 
sorted data lists. The quick sort is known as 
the best, and fastest sorting algorithm of $O(N\  log\  N)$ performance. 
However, the quick sort does falter on 
already sorted data lists, degenerating into a O(n2) time complexity 
algorithm. Thus, the bubble sort is best 
when the quick sort is at its worst, and vice-versa. Again the extremes 
of algorithmic performance in pre-
existing algorithms have been selected for comparison and contrast with 
the hash sort algorithm. 

\subsection{Test Cases}

\indent\indent

Testing on data size looks at increasing data sizes and the rate of 
growth in the time complexity of the 
algorithm. Testing on data size is on "complete" lists of data elements, 
lists which are fully sorted or are 
unsorted. These types of  "complete" data test cases form the extremes of 
the other variations in the data test 
case. The sorted lists are either fully sorted in ascending or descending 
fashion, and the unsorted list is the 
median between the two extremes. Size testing looks at the effects of 
increasing sizes on these three types of 
"complete" sorted data test cases.

\subsection{Test Program}

\indent\indent

The test program has to handle to important issues dealing with the 
program code, and the test platform. 
The test platform, or the computer system the code is run on, the test 
program must not be biased by any 
hardware or platform architecture enhancements. So a general-purpose 
computer, without any specific 
enhancements or features is used. For further platform independence, 
different computers should be used, of 
different sizes and performance to be sure of the independence of the 
results obtained. 

\vspace{3 mm}

The code must be written in a  "neutral" computer language to avoid any 
esoteric features of a language, 
and be readable. Similar to platform independence, the tests must be 
independent of the language implementing 
the test programs. Any programming language features which give a biased 
advantage in the generated code 
being optimized or better for is to be avoided. 

\vspace{3 mm}

The code must be readable, so a programming language which is expressive 
along with good programming 
style must be used. Unreadable code will hamper duplication of results by 
others and cast doubt on the efficacy 
and credibility of the tests. The code must be portable, so that 
independent platform testing can be conducted, 
and require minimum effort to re-write and re-work the code for continued 
testing. 
\vspace{3 mm}
The test case generator and the test sort programs are integrated 
together in one program. The test case 
generator generates the particular test case in internal storage within 
the program, and the hash sort, bubble sort, 
and quick sort all process the data. The time to process the data is 
noted and written to an output data file. The 
time in microseconds, and the size of the data set is recorded by the 
program.

\subsubsection{Test Platform Specifications}

\vspace{3 mm}

\noindent
The table summarizes the different computer systems which were used to 
test the hash sort.

\vspace{3 mm}

%table of test platforms

\begin{quote}
\begin{tabular}{|c|c|c|c|}
\hline
Processor: & Sparc & Sparc & Cyrix i686-P200 \\
\hline
Hardware: & sun4u & sun4m & vt 5099A \\
\hline
Operating System: & SunOS 5.5.1 & SunOS 5.5.1 & Linux 2.0.33 \\
\hline
Processors: & 12 & 4 & 1\\
\hline
Total Memory: & 3072 Megabytes & 256 Megabytes & 64 Megabytes \\
\hline
GNU gcc Version: & 2.7.2.3 & 2.7.2.3 & 2.7.33 \\
\hline
\end{tabular}
\end{quote}

\vspace{ 3mm}

The original Pascal implementations were converted to C using the the p2c 
converter. The generated C 
code was hand optimized to eliminate inefficiencies and to follow strict 
ANSI C compliance. The code was 
modified for the UNIX system environment, in particular the system timing 
calls. Some of the routines such as 
write\_matrix and write\_list were originally used to check that the 
output was a sorted list or array -- verifying 
the algorithms were working. Once this had been checked, the call to the 
procedure was commented out, as the 
actual test program dumped its output to a data file. These routines in 
the final run for testing timing are still 
present in the source code.

\section{Analysis of Results}

\subsection{Test Expectations}

\indent\indent

The expectations of the tests are for the quick sort and bubble sort to 
perform as outlined. Both algorithms 
have been extensively studied over time, so any change in these 
algorithm's performance and behavior would 
be a shocking surprise. The hash sort is expected to remain linear, even 
as the bubble sort and quick sort fall 
apart in their worst case scenarios. The hash sort will falter, but will 
not degenerate as badly as the bubble sort 
and quick sort do in the extreme cases. Nor is the hash sort expected to 
have any special cases of superior 
performance. 

\vspace{3 mm}

The hash sort will have linear time complexity performance which will 
have a Big-Oh of $O(N)$ in relation to 
the test case size. This consistency and stability of the algorithm 
through different cases where the comparison 
algorithms degenerate or accelerate performance is expected to be 
verified in the tests. The worst case of the 
algorithm performance will be a constant multiple of linear time $c \cdot 
N$, but it will remain linear time-complexity.

\subsection{Test Case Results}

\indent\indent

The table of the run-time performance of each algorithm in the appendix 
shows an increasing time with 
increasing size of the data set. The bubble sort "explodes" as expected, 
and soon is surpassed by the hash sort 
and quick sort. Both the hash sort time and the quick sort time increase, 
but what is of interest is the ratio of the 
two. The hash sort does not immediately surpass the quick sort, in fact 
it is not until the data set  $N \geq 146$ that 
this occurs. A steady trend of the hash sort gaining occurs up to data 
set $N < 120$. When the data set is in the 
range   $( 120 \leq N < 146 )$ the hash sort to quick sort performance 
ratio occasionally is less than one. The 
performance fluctuates until data set $N > 145$, when the ratio of the 
hash sort to the quick sort $ < 1.0$, meaning the 
hash sort is performing faster than the quick sort. A decreasing ratio 
was expected, but the hash sort did not 
immediately make gains on the quick sort until the data set was much 
larger.

\section{Conclusions}

\subsection{Testing Conclusions}

\indent\indent

The hash sort performed as expected, as did the bubble sort and the quick 
sort algorithms. What was 
surprising initially is that the hash sort does not have a performance 
lead over the quick sort. The bubble sort 
was soon surpassed by the hash sort as expected, but the hash sort did 
not exceed the performance of the quick 
sort as it should have theoretically. 

\vspace{3 mm}

The reason for this seemingly strange performance of the algorithm is the 
theoretical consideration that the 
hash sort and the quick sort operate using the same types of operations 
with the same underlying machine code 
clock cycles. This of course is not correct, the quick sort is using a 
compare based machine instruction, whereas 
the hash sort is using an integer divide machine instruction.

\vspace{3 mm}

A comparison of the clock cycle times for the compare and integer divide 
instructions on a Intel 486 
processor gives some indication of this (Brey pp. 723 - 729).

\vspace{3 mm}

%table for the op-codes

\begin{quote}
\centering
\begin{tabular}{|c|c|c|}
\hline
Opcode & Addressing Mode & Clock Cycles \\
\hline\hline
CMP & register-register & 1 \\
\hline
CMP & memory-register & 3 \\
\hline
DIV & register & 40 \\
\hline
DIV & memory   & 40 \\
\hline
IDIV & register & 43 \\
\hline
IDIV & memory  & 44 \\
\hline
\end{tabular}
\end{quote}

\vspace{3 mm}

The ratio of a compare instruction to a integer divide is 1:42 for 
register to register based machine 
instructions, and 1:21 for memory based machine instruction. The summary 
of this is that the quick sort and the 
hash sort are utilizing different machine instructions to implement the 
algorithm, and theoretically they are 
equivalent at an abstract level. At the implementation level, this is not 
a valid assumption to make. A simple 
analysis of the algorithms in this context shows this.

%Ratio of two algorithms

\vspace{3 mm}
\noindent

\[
\begin{array}{lcl}
F(Quick\ sort) & > & F(Hash\ sort)\\

c_2 \cdot N \cdot log_2 N &\ \ \ \ \  >\ \ \ \ \ \ &  c_1 \cdot 2 \cdot N 
\mbox{ dividing through by }N \\

c_2 \cdot log_2 N & > &  c_1 \cdot 2\ \mbox{ dividing through by }c_2 \\

log_2 N \ & > &  c_1 \cdot 2\ \mbox{ substituting } N = \frac{146^2}{c_2} \\

log_2(146^2 ) & > &  \frac{c_1 \cdot 2}{c_2} \mbox{ move square down} \\

2 \cdot log_2 (146) & > &  \frac{c_1 \cdot 2}{c_2}\ \mbox{ divide through 
by } 2 \\
				    
log_2 (146 ) & > &  \frac{c_1}{c_2}\ \mbox{ evaluate the log} \\

14.38 & > &  \frac{c_1}{c_2}
\end{array}\]

\vspace{3 mm}

This rough analysis is an approximation of the arithmetic instruction 
time to comparison instruction time. It 
does not directly correspond to the ratios of clock cycles for the Intel 
instructions. There are other factors, such 
as compiler optimization, and memory manipulation overhead to be 
considered. But this rough analysis does 
give some insight into the "critical mass" that the hash sort must reach 
before it can exceed a comparison based 
algorithm -- in this case the quick sort.

\subsection{Evaluation Criteria}
\noindent
The three algorithms used are to be evaluated on the basis of the 
following criteria:

\vspace{3 mm}

\begin{itemize}

\item coding complexity -- how difficult or easy is it to code the algorithm?

\vspace{3 mm}

\item time complexity -- how fast is the algorithm in performance?

\vspace{3 mm}

\item space complexity -- how efficient is the algorithm in using memory?

\vspace{3 mm}

\end{itemize} 

To rank the algorithm on each of the criteria, the simple expedient of 
following a grading system is used. 
Rankings are: excellent, good, fair, poor, bad. A comment as to the 
ranking and why it was judged to rate such a 
score is given. 

\subsection{Evaluation of the Algorithm}
\noindent
The hash sort has the following rankings:

\begin{itemize}

\vspace{3 mm}

\item coding complexity:  excellent; the hash sort is a short algorithm 
to encode, uses only 	a simple iterative loop 
and array manipulation.

\vspace{3 mm}

\item time complexity:  good; the hash sort is linear in theory, but in 
practice must reach a certain critical mass, so 
its overall time complexity is more degenerate than the theory behind the 
algorithm would indicate.

\vspace{3 mm}

\item space complexity:  fair; the hash sort has some overhead in the 
algorithm, but of which most are counters 
which are incremented as the algorithm progresses. But the algorithm is 
fitting a n-dimensional array onto a 
one-dimensional memory so this is somewhat awkward and introduces memory 
manipulation complexity 
overhead. 

\vspace{3 mm}

\end{itemize}

\subsection{Comparison to Test Algorithms}

%table of comparison to test algorithms

\begin{quote}
\centering
\begin{tabular}{|c|c|c|c|}
\hline
Criteria & Hash Sort & Quick Sort & Bubble Sort \\
\hline\hline
Coding Complexity & Excellent & Fair & Excellent \\
\hline
Time Complexity & Excellent & Good & Poor \\
\hline
Space Complexity & Fair & Good & Good \\
\hline
\end{tabular}
\end{quote}

\vspace{3 mm}

\subsection{Summary of Sort Algorithms}

\indent\indent

The hash sort compared to the bubble sort is similar in coding 
complexity, very short code which is simple 
and easy to follow. However, it is much faster than the bubble sort, 
although it is not as efficient in using 
memory. The bubble sort is much more efficient in using memory, but it 
does have much data movement as it 
sorts the data. 

\vspace{3 mm}

The quick sort is much more complex to code, even the recursive version. 
A non-recursive 
version of the quick sort becomes quite complex and ugly in coding. The 
quick sort does outperform the hash 
sort initially, but it is still a linearithmic algorithm, and does have a 
degenerate worst case, although it is rare in 
practice. The quick sort is good at managing memory, but it is a 
recursive algorithm, which has somewhat more 
overhead than an iterative algorithm. Because of its partitioning and 
sub-dividing approach, the amount of data 
movement is less than with bubble sort.

\subsection{Further Work}

\indent\indent

The initial development and investigation into the properties, 
performance, and promise of the hash sort 
seem thorough, there is still other avenues of research. Further research 
on the hash sort can be done in terms of 
investigating a recursive version of the algorithm, seeing if it can be 
parallelized, remedying the problem of 
data sparsity, and finding other mapping functions. 

\subsubsection{Recursive implementation:}

\indent\indent

The hash sort has been implemented as an iterative algorithm, but some 
sorts, such as the quick sort, are 
inherently recursive. A recursive version of the hash sort may be 
possible, but it may not translate well into a 
recursive definition or implementation. The possibilities of the hash 
sort being implemented a smaller 
mappings of the same data set is an interesting possibility. 
	
\subsubsection{Parallelization of algorithm:}

\indent\indent

The hash sort algorithm has been implemented as a serial algorithm on a 
uniprocessor system, although two 
of the systems it was tested on had multiple processors. It is possible 
that the hash sort could be parallelized to 
increase its performance beyond linear time. The tradeoffs of such a 
parallel algorithm, and the issues involved 
are another possibility for research. 

\subsubsection{Sparsity of data:}

\indent\indent

The problem of sparse data within the range of the hash sort is a problem 
with the algorithm. There may be 
ways of handling sparse data so that space can be utilized more 
efficiently than the implemented algorithm here 
does. If the problem with data sparsity can be resolved, then the hash 
sort could become a magnitude more 
useable in applications. 

\subsubsection{Other mapping functions:}

\indent\indent

The heart of the hash sort is the super-hash functions, which are perfect 
hashing which preserve ordering. 
This mapping function used the classical hash function along with a mash 
function to hash magnitude. Other 
mapping functions which have the same properties would be very 
interesting functions in themselves, but could 
be the basis for another variant of the hash sort. A less system complex 
mapping function that did not use the 
time consuming div and mod operations would be an obvious improvement in 
the hash sort. 

\subsubsection{Machine code optimization:}

\indent\indent

The difference in the machine instructions used to implement the 
underlying algorithms makes the hash sort 
need a "critical mass" to reach a sufficient size before it meets its 
theoretical performance expectations. 
Optimizations of the machine generated code to reduce the size of the 
critical mass required are an interesting 
possibility. Such optimizations could be compiler-based, or strictly upon 
the properties of the hash sort 
algorithm. 

\subsubsection{Alternative data structure:}

\indent\indent

The current version of the hash sort uses multiple $N$ square matrices 
for the data structure being 
mapped to by the super-hash function. One possibility to be explored is 
using non-square matrices, and
possible other data structures that are not square or rectangular. A 
change in the data structure mapped
to reflects a change in the mapping function, as the mapping function is 
determined by the data structure
being mapped to by the super-hash function. 

\section{Application}

\subsection{Criteria for using Hash Sort}

\indent\indent

The hash sort, although a general-purpose algorithm, is not an algorithm 
that meets all needs for 
applications. Because of its properties and features, there are several 
criteria which guide using it. These criteria 
are:

\begin{itemize}

\vspace{3 mm}

\item data set is within a known range 

\vspace{3 mm}

\item data set is numeric or inherently manipulatable as numbers

\vspace{3 mm}

\item the data needs to be ordered quickly then accessed frequently large 
data set with heavy density within the data range

\vspace{3 mm}

\end{itemize}

\subsection{Applications for Hash Sort}

\indent\indent

There applications suited for the hash sort can be surmised from the 
criteria mentioned previously, but some 
applications include:

\begin{itemize}

\vspace{3 mm}

\item database systems -- for high speed lookup and retrieval

\vspace{3 mm}

\item data mining -- organizing and searching enormous quantities of data

\vspace{3 mm}

\item operating system -- file organization and access 

\vspace{3 mm}

\end{itemize}

\subsection{Example Application with Hash Sort}

\indent\indent

One interesting application of the hash sort is with  data 
communications. This example involves data 
communication on a X.25 packet switched network. In this method of data 
communications, data is 
disassembled into a packet, which is then sent through the network. At 
the reception point, the packets are 
reassembled into the original data. One problem with  the X.25 packet 
switched network, the packets are out of 
order when received from the sender, or "...may arrive at the destination 
node out of sequence." (Stallings 1985 
p. 249) That is, data can arrive in sequence as sent, but it may not. 
This is because of "...datagrams are 
considered to be self-contained, the network makes no attempt to check or 
preserve entry sequence." (Rosner 
1982 p. 117)

\vspace{3 mm}

There is a need to sort the packets once they are received. This is not 
so straightforward, because there is 
the offhand chance that the packets may be in order, so sorting may be 
unnecessary. Using a quick sort on a 
sorted sequence would be the worst-case degenerate example, the quick 
sort would become an $O(N^2)$ algorithm 
-- as bad as bubble sort. 

\vspace{3 mm}

Even worse, all the packets must be received before they can be sorted, 
if sorting is required. To use a quick 
sort for this application, a separate algorithm to check if the packets 
are already sorted is required, then if not, 
the quick sort algorithm is used. 

\vspace{3 mm}

Overall this is adding addition time overhead in terms of:

\vspace{3 mm}

\[ Total\  time\  =\  Time(receive)\  +\  Time(check\  sorted)\  +\  
Time(quick\  sort) \]

\[ Total\  time\  =\  Time(r)\  +\  Time(N)\  +\  Time(\  N  \cdot log N) 
\] 

$N$ is the size of the packets received, and to check that the packets 
are in order requires a linear search $O(N)$ 
to examine all packets, although not when they are out of order. 

\vspace{3 mm}

The hash sort is a better algorithm to use, although the packets would 
have to be stored in a matrix structure 
rather than a linear list. A check for the packets being received in 
order is unnecessary, because the hash sort 
will not degenerate on sorted data -- it will be an extreme form of 
hysteresis, but the hash sort remains linear in 
such cases. 

\vspace{3 mm}

More interesting is that the hash sort does not have to wait for the last 
packet to be received -- it can begin 
to sort the packets as they are received. The hash sort maps the packets 
to where each uniquely belongs. Since 
no two packets share the same location, they are unique. So the hash sort 
can uniquely determine from the 
packet itself where it belongs in the matrix. As packets are received, 
they are placed into their appropriate 
location. By the time of the last packet arrival, all the data is in 
place. 

\vspace{3 mm}

The hash sort may be slower than the data reception time, in which case 
the algorithm will continue to sort 
the packets as they are stored. The total time in this case would be:

\vspace{3 mm}

\[ Total\  time\  =\  Time(receive)\  +\  Time(hash\  sort) \]

\[ Total\  time\  =\  Time(r)\  +\  Time(N) \]

\vspace{3 mm}

This time is the same as if the data were received in the quick sort 
implementation already in order, and the 
algorithm to check that the data is in order is used. This example shows 
the power of data independence in the 
hash sort algorithm -- a data element does not depend on its predecessor 
or successor for its location, it is 
determined uniquely by the data element itself. 

\vspace{3 mm}

This property of data independence, and the speed of the hash 
sort and its robustness in the degenerate case, make such applications 
where it is applicable to use it. One 
drawback however, is that the size of the packets sent must be sufficient 
enough to achieve the "critical mass" 
needed by the hash sort. 

\section{Mathematical Proofs}

\subsection{Proof of Injective Mapping Function}
\[ \mbox{\em Proof for a general super-hash function with a n-tuple of 2}\]

\noindent
\underline{Theorem:}

\vspace{3 mm}

Given a set $S$ of unique integers in $Z$ such that 
$S = \{ v_1 \not= v_2 \ldots v_{n-1} \not= v_n \} $, and given a function
$F(x)$ such that:

%Define the function and identify its component parts
\[
F(x) \equiv \left\{ \begin{array}{ll}
p & = x\ \mathbf{div}\ n;\mbox{\ \ \ \ \ \ \ \ Mash function }\\
q & = x\ \mathbf{mod}\ n;\mbox{\ \ \ \ \ \ Hash function }\\
\end{array} \right.
\]

\vspace{3 mm}
Where $n > 0$, then the resulting image set $M$ of ordinal pairs formed is
$M = {(p_1,q_1),(p_2,q_2), \ldots ,(p_{n-1},q_{n-1}),(p_n,q_n)}$ from the 
pre-image
set $S$ is an injective mapping $F \colon S \rightarrow M$.

\vspace{3 mm}
\noindent
\underline{Proof:}
\vspace{3 mm}

Proof by Contradiction: Proceed with the assumption that given $S$ as 
described before that
$\forall S$, $\exists (x,y) \in S$ such that $F \colon X = F \colon Y$ 
when $x\ \not= \ y$, or 
that the function $F\colon S \rightarrow M$ is a non-injective mapping.

The definition of injective (one-to-one) is "A function $f$ is one-to-one 
if and only
if $f(x) = f(y)$ implies that $x = y$ for all $x$ and $y$ in the domain 
of the function."

\[
[ \forall (x,y) \in F \colon X \rightarrow Y, \mbox{where } (f(x) = f(y) 
) \supset (x = y)
  \mbox{ is injective}
]
\]

By taking the contrapositive of this definition, it is restated as "A 
function $f$ is one-to-one
if and only if $f(x) = f(y)$ whenever $x \not= y$ in the domain of the 
function." (Rosen 1991, p.57)

\[
[ \forall (x,y) \in F \colon X \rightarrow Y, \mbox{where } \neg(x = y) 
\supset 
  \neg ( f(x) = f(y) ) \mbox{ is injective}
]
\]
\noindent
rewrite as

\[
[ \forall (x,y) \in F \colon X \rightarrow Y, \mbox{where } (f(x) \not = 
f(y) ) 
  \supset (x \not= y)  \mbox{ is injective}
]
\]

By taking the contrapositive of this definition, it can be restated as "A 
function $f$ is one-to-one
if and only if $f(x) \not= f(y)$ whenever $x \not= y$. (Rosen 1991, p.57)

\vspace{3 mm}

Using the contrapositive of the definition of an injective function, it 
is readily clear that the mapping
$F \colon S \rightarrow M$ is not injective if there are at least two 
integers $i_1$ and $i_2$ such that by
the mapping function $F$, $(p_1,q_1) = (p_2,q_2)$ in $M$. This is assumed 
to be true, as a non-injective 
mapping function.

\vspace{3 mm}

By the definition of $S$, all the integers are unique in $Z$, so the 
integers have the property that for any
two integers $x$,$x' \in S$ so that $x \not= x'$.

\noindent
Take the case for x:

\[ \begin{array}{ll}

p_x = x\ div\ n & q_x = x\ mod\ n \\
p_x = d_x \mbox{ where } d_x \geq 0\ \ \ \ \ \ \ \ \ \  & q_x = m_x 
\mbox{ where } 0 \leq m_x < n

\end{array}\]

\noindent
Take $x' = x + c$ where $0 < c < n$:

\[ \begin{array}{ll}
p_x = x'\ div\ n & q_x = x'\ mod\ n \\

p_x = (x + c)\ div\ n & q_x = (x + c)\ mod\ n\\

p_x = (x\ div\ n) + (c\ div\ n) & q_x = (x\ mod\ n) + (c\ mod\ n)\\

p_x = d_x + k \mbox{ where } d_k, k \geq 0\ \ \ \ \ \ \ \ \ \  & q_x = 
m_x + c \mbox{ where } 0 \leq m_x < n

\end{array}\]

\noindent
so then that for $x = x'$ that $F(x) \not= F(x')$. But this is for $x' = 
x+c$ where $0 < c < n$.
Take the case when $c \geq n$. Using the same definition for $x$, take 
$x' = x+c$ where $c = n$.

\[ \begin{array}{ll}

p_x = x'\ div\ n & q_x = x'\ mod\ n \\

p_x = (x + n)\ div\ n & q_x = (x + n)\ mod\ n\\

p_x = (x\ div\ n) + (n\ div\ n) & q_x = (x\ mod\ n) + (n\ mod\ n)\\

p_x = d_x + 1 \mbox{ where } d_x \geq 0\ \ \ \ \ \ \ \ \ \  & q_x = m_x + 
0 \mbox{ where } 0 \leq m_x < n\\

p_x = d_x + 1 \mbox{ where } d_x \geq 0 & q_x = m_x \mbox{ where } 0 \leq 
m_x < n
\end{array}\]

\noindent
so then that for $x \not= x'$ that $F(x) \not= F(x')$.

\vspace{ 3 mm}

For each case of the form $x' = x + c$, where $c < n$ and $c \geq n$, if 
$x = x'$ then $F(x) = F(x')$
which contradicts with the original assumption that $\exists (x,y) \in S$ 
so that $F \colon X = F \colon Y$
when $x \not= y$.

\vspace{3 mm}

So the occurence of $d_x = d_x'$ and $m_x = m_x'$ such that $x = x'$ is 
never true and will never occur, which 
is the only possibility for the mapping $F \colon S \rightarrow M$ to be 
non-injective.

\vspace{3 mm}

Thus it follows that no two ordinal pairs in $M$ formed from any two 
integers in $S$ by $F$ would ever be equal.
Since this is the only counter-example for the definition of $F(x)$ to be 
an inejective mapping, then the only
conclusion is that $F \colon S \rightarrow M$ must be an injective 
mapping under $F(x)$.

\vspace{ 3 mm}

\noindent
\textbf{Q.E.D.}

\subsection{Proof of Injective Multiple Mappings}
\[ \mbox{\em Proof using general super-hash function}\]

\noindent
\underline{Theorem:}

\vspace {3 mm}

For an injective mapping $F \colon S \rightarrow M$, using a set of 
mapping value $n \in N$, the set of 
unique mappings values used in the super-hash function, that a function 
composed of multiple sub-mappings
of $F$ with $n \in N$ is a complete mapping function overall. Therefore 
such a constructed function would
be an injective mapping.

\vspace{3 mm}

\noindent
\underline{Proof:}

\vspace {3 mm}

Proof by Demonstration: Define a function $G$, which non-redundantly 
selects $n_x \in N$, where $x \leq \parallel N 
\parallel$, the cardinality of the set of elements in $N$. Therefore, 
each selected element is guaranteed by the 
definition to be unique from the predecessor or successor elements, so 
that no element once selected as a mapping value 
will be selected again. So the function $G$ selects an element once and 
only once from $N$.

\vspace{3 mm}

The selection function $G$ is an injective function by its definition. 
For each $n_k \in N$ there is an integer from $1 \leq x
\leq 
\parallel N \parallel$ which is uniquely associated with the particular 
element selected $n$. The definition of an injective
function
is "A function $f$ is one-to-one if and only if $f(x)=f(y)$ implies that 
$x=y$ for all $x$ and $y$ in the domain of the
function."

\vspace{3 mm}

For the particular selection or iteration of $G$ which is $k$, there is a 
unique element $n_k$, selected only once by $G$ from
$N$. 
Thus, iterations $k-1$, $k$, and $k+1$ will not select the same element, 
so $G(k) = n_k$, so that $k = n_k$ for the set $N$
under 
the selection function $G$. Hence the definition and construction of $G$ 
make an injective function.

\vspace{3 mm}

The mapping function $F\colon S \rightarrow M$ has been proven to be an 
injective mapping function previously for $n$, where
$n$ is 
the mapping value used in the mash and hash sub-functions. Composing a 
new function using the mapping function $F$ with the
selection
function $G$, the resulting composed function $H$ is an injective mapping.

\vspace{3 mm}

For $F \colon S$ is using a mapping value $n$, selected from $N$. So $F 
\colon S$ could be rewritten as:

\[ F(n) \colon S \]

going one step further,

\[ F(G) \colon S \]

and moreover since $G$ uses $N$ as its domain, then

\[ F(G:N):S \]

The notation is rapidly becoming cumbersome, so in general using a 
mapping function $F$ along with a selection function $G$,
we have a
function $H = F \circ G$, meaning the function $H$ is a composition of 
the functions $F$ and $G$. With this in mind, the proof
that $H$
is a injective function is very straightforward, and involves the 
property that $F$ and $G$ are injective functions.
Therefore, the composite
function $H$ is also an injective function. 

\vspace{ 3 mm}

\noindent
Proof of composite of two injective functions is also injective is:

\vspace{3 mm}

If $f:X \rightarrow Y$ and $g:Y \rightarrow Z$ are injections (injective 
functions), then so is the composite of $g \circ f:X
\rightarrow Z$.
Suppose $gf(x) = gf(x')$. Since $g(f(x)) = g(f(x'))$, and $g$ is an 
injection, we must have $f(x) = f(x')$, and since $f$ is
an injection, $x = x'$.
Hence $g\circ f$ is an injection. (Biggs 1985, p. 31)

\vspace{3 mm}

For the function $F$, is uses a particular mapping value $n$, where $G$ 
selects $n \in N$. So it follows that $G(k):N
\rightarrow n$, for a particular
iteration or selection $k$. $F(n):S \rightarrow M$ has already been 
proved as an injective mapping by and of itself. For
multiple iterations of $k$,
$F(G(k):N):S \rightarrow M$ is an injective mapping.

\vspace{3 mm}

The power of $F$ to map from one mapping instance, to muliple $k$ 
iterations is expanded using a mapping value set $N$, and a
function to uniquely
select particular elements as mapping values $G$. Hence the injective 
mapping property of $F$ composed with $G$ has been
generalized into multiple 
mappings.

\vspace{ 3 mm}

\noindent
\textbf{Q.E.D}

\end{document}